\documentclass{iaus}
\usepackage{graphics,amsmath}

\title[Turbulence and Dynamo Interlinks] 
{Turbulence and Dynamo Interlinks}

\author[de Gouveia dal Pino et al.\ ]   
{E. M. de Gouveia Dal Pino$^1$,
\ R. Santos-Lima$^1$,
\ G. Kowal$^1$,
\break \ and D. Falceta-Gon\c calves$^{2}$,
}

\affiliation{$^1$ IAG, Universidade de S\~ao Paulo, Rua do Mat\~ao 1226,  S\~ao Paulo 05508-090, Brazil
\\ email: {\tt dalpino@astro.iag.usp.br} \\[\affilskip]
$^2$ EACH, Universidade de S\~ao Paulo, Rua Arlindo Bettio 1000,  S\~ao Paulo 03828-000, Brazil
}

\pubyear{2013}
\volume{294}  
\pagerange{}
 \jname{Solar and Astrophysical Dynamos and Magnetic Activity
}
\editors{A. G. Kosovichev, E. M. de Gouveia Dal Pino, Y.Yan, eds.}
\begin{document}

\maketitle

\begin{abstract}
The role of  turbulence in astrophysical environments and its interplay with magnetic fields is still highly debated. In this lecture, we will discuss this issue in the framework of dynamo processes.
We will first present a very  brief summary  of turbulent dynamo theories, then  will focus on small scale turbulent dynamos and their particular relevance on the origin and maintenance of magnetic fields  in the intra-cluster media (ICM) of galaxies. In these environments, the very low density of the flow requires a collisionless-MHD treatment. We will show the implications of this approach in the turbulent amplification of the magnetic fields in these environments.  To finalize, we will also briefly address the connection between MHD turbulence and $fast$ magnetic reconnection and its possible implications in the diffusion of magnetic flux in the dynamo process.

\end{abstract}

\firstsection 
\section{Introduction}
Magnetic fields and turbulence are  ubiquitous in  the Universe. Their interplay is currently  a matter of intense study and debate particularly because magnetohydrodynamical (MHD) turbulence is still a theory in development.

Conceptually, turbulence is the state of an irregular fluid subject to energy injection at large length scales and energy dissipation at short length scales (e.g. Brandenburg et al. 2012). The ratio between both the injection and the   dissipation scales   can be quantified by the Reynolds number.
 This is  typically  very high
owing  to  the large  scales  of  the astrophysical  fluids  compared  to the  small  dissipative  scales.

One  of  the  central
questions of MHD dynamics is how initially unmagnetized  well-conductive fluids  generate their  own  magnetic field, namely how dynamo action takes place.

Dynamos involve  the conversion of  kinetic energy  into magnetic energy and turbulence (when present) is believed to play a  key role on this process: (i) through the amplification of seed magnetic fields (MFs) by stretching (or local shear); (ii) helping to maintain developed MFs; and (iii) through the dissipation of small scale MFs by turbulent diffusion (the so called beta effect) and fast reconnection.

There are several recent excellent reviews both on dynamos (see e.g., Brandenburg \& Subramanian 2005; Schekochihin et al. 2004; Schekochihin et al. 2007;
Brandenburg et al. 2012; Beresniak 2012) and on MHD turbulence theory (e.g., Eyink 2011;
 Verma 2004; Eyink et al. 2011; Lazarian 2011, see also B. Burkhart these Proceedings; to mention just a few).
  In this lecture we will first present a very  brief summary  of turbulent dynamo theories, then  will focus on small scale turbulent dynamos and their role on the origin and maintenance of magnetic fields  in the intra-cluster media. In this framework, we will discuss, in particular, the importance in taking into account the effects of a collisionless-MHD approximation. To finalize, we will also briefly address the connection between MHD turbulence and $fast$ magnetic reconnection and its possible implications on the diffusion of magnetic flux in the dynamo context. (For a complete review on the current knowledge of both large and small scale turbulent dynamos we also refer to other chapters in these Proceedings; see, e.g., A. Brandenburg; E. Vishniac, and J. Schober압  chapters.)

\section{
 Turbulent dynamos
}

  Turbulent dynamos are generally subdivided into  large-scale dynamos
and small-scale or fluctuation dynamos depending on
whether  magnetic  fields  are amplified  on  scales smaller
or  larger  than  turbulence  outer  scales, respectively (e.g., Brandenburg et al. 2012; Beresniak 2012).

 Large-scale (LSDs) dynamos  are generated when statistical symmetries of the turbulence are broken by large-scale asymmetries of the system such as stratification, differential rotation and shear (Vishniac \& Cho 2001; K\"apyl\"a et al 2008, Beresniak 2012).
Turbulent flows possessing perfect statistical  isotropy cannot  generate large-scale magnetic fields.
The so called  twist-stretch-fold  mechanism   introduced by Vainshtein \&
Zeldovich (1972) was conceived for  generating   large-scale fields.

 Large-scale dynamos can be also referred to as mean-field
dynamos since the field evolution can be obtained from the mean field theory, namely, by averaging the governing equations, particularly
the induction equation.  LSDs can be  excited by helical turbulence and are expected to generate magnetic fields in astrophysical sources such as the sun and stars, accretion disks, and disk galaxies.

Small-scale dynamos (SSDs) on the other hand, can be  excited by non-helical turbulence and are believed to be a key dynamo process, for instance,  in the intra-cluster medium of galaxies (Subramanian et al. 2006).
First predicted by Batchelor (1950) and others (Biermann \& Schl\"uter 1951; Elsasser 1956),
 SSDs can work, in principle, under fully isotropic conditions and are important in cosmic objects because they are generic to any random flow with sufficiently conducting plasma.

 The currently accepted theoretical grounds for SSDs were provided by Kazantsev (1967).
 SSDs  are faster than LSDs in most astrophysical environments and the magnetic
energy grows in the beginning  exponentially  upto equipartition with the kinetic energy at  the eddy turnover timescale of the smallest eddies (Subramanian 1998).
Later, it grows  linearly at the turnover timescale of the larger eddies (Beresnyak 2012),
with the largest scales of the resulting field being a fraction
of the outer scale of the turbulence.
Both time scales are typically much shorter than the age of the system.
For instance, in the case  of galaxy clusters, the typical scale and velocity of the turbulent eddies are  around 100 kpc and 100 km/s, respectively, implying  a growth time $\sim 10^8$ yr which is much smaller than the typical ages of such systems. This means that
SSDs should operate and are actually crucial to explaining the observed magnetic
fields  on the scales of tens of kiloparsecs in the intra-cluster media. Besides, according to Subramanian et al. (2006), it would be hard to explain magnetic
fields  on larger scales is such environments because the conditions for LSD action are probably
absent.

Small-scale dynamos are currently  also invoked to describe the small-scale magnetic
field at the solar surface (see Brandenburg et al. 2012 and references therein).
As remarked by Brandenburg et al. (2012), in many contexts both SSDs  and LSDs should  go together and it is not clear whether one can  distinguish 
a small-scale field from an SSD from that associated with the fluctuations  that
are inherent to any LSD and that can be caused by tangling and amplification
of the large-scale field.
In fact, even in an SSD, after the initial fast growth,
the turbulent fields  can be slowly ordered by mean-field dynamo process (if the conditions in the system allow for it),
with turbulent diffusivity and magnetic reconnection diffusion of MHD turbulence playing an essential role (see also Section 4).

\subsection{Small scale versus large scale turbulent dynamos}

Figure 1 illustrates the growth of the magnetic field in  a large scale turbulent dynamo. In this case, a  3D MHD numerical simulation of a rotating turbulent convection system
was performed (Guerrero \& de Gouveia Dal Pino 2010).  Rotating convection is a natural scenario for the study of helical turbulent dynamo action and the generation of  large scale magnetic fields, as  those  observed in the sun (see also  Cattaneo \& Hughes 2006, K\"apyl\"a et al. 2008). Figure 1 shows the magnetic field structure after the system evolved for about 40 turnover times of the turbulence. The presence of rotation allows the exponential amplification of an initial  seed field to a large scale
magnetic field to  near equipartition values with the convective motions (Guerrero \& de Gouveia Dal Pino 2010).

\begin{figure}
\centering
\resizebox{6.3cm}{!}{\includegraphics{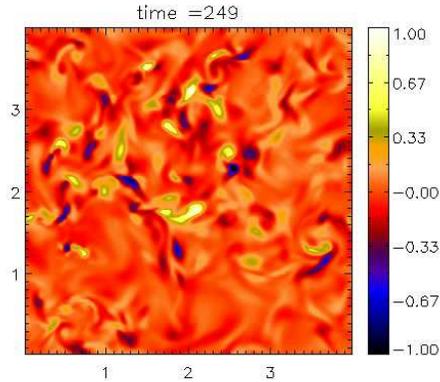} }
\caption[]{Central slice of a 3D box with open boundary conditions showing magnetic field amplification in a rotating  convective turbulent system. It was started   with a
polytropic gas in hydrostatic equilibrium bounded  by a stable
overshoot layer at the bottom and a convectively unstable layer at the top of the computational
domain. This hydrodynamic system evolved  until a steady state helical turbulent regime developed
and then a seed magnetic field was introduced and the  system evolved
for about more  40  turnover times. Magnetic field structures amplified by the LSD action (through the alpha effect) are  detected (see also Guerrero \& de Gouveia Dal Pino 2010). }
\label{fig1}
\end{figure}

Another comprehensive  example showing  how SSDs and LSDs should  behave is given in Figure 1 of Brandenburg et al. (2012) which compares the computed spectrum of the magnetic and kinetic energies, $E_M (k$ )
and $E_K (k)$ (where $k=2\pi/l$ is the turbulence wavenumber at a given scale $l$)), respectively, for two  systems containing  seed magnetic fields, one with helical turbulence injected (i.e., with non-null  kinetic helicity, $\mathbf{v . (\nabla \times v}) \neq 0$, where $\mathbf{v}$ is the turbulent velocity),  and the other one with non-helical turbulence injected.  The first case allows for an LSD, while the second case for an SSD action.
In both cases it is found that in  the early times the dynamo evolution is
quite similar: both  have a $k^{3/2}$  power spectrum at small scales, as
predicted by Kazantsev (1967) (see also Kulsrud \& Anderson 1992)  for
SSDs. The growth of the magnetic energy saturates in the SSD case when it reaches equipartition with the kinetic energy.  However, in the case with  helical turbulence (i.e., the LSD case), at late times there is  the development of an inverse cascade with energy deposition at scales $l$ larger than the turbulent injection scale ($l>l_{inj}$). This is  due to the presence of  the kinetic helicity (it is the so called alpha effect in LSDs, which is discussed in more detail elsewhere in these Proceedings; see e.g., Brandenburg and Vishniac's chapters).

\subsection{Turbulent dynamo characteristic numbers and the saturation condition of the magnetic fields in SSDs}

In the study described in the previous paragraph, the adopted magnetic Prandtl number in the simulated systems was $Pr_M=1$.
This number is given by the ratio between the magnetic Reynolds number ($Re_M$) and the Reynolds number $Re$. In a  turbulent flow, $Re=v_{rms}/k_{inj} \nu$, while  $Re_M=v_{rms}/k_{inj} \eta$, where $v_{rms}$ is the root mean square of the turbulent velocity, $\nu$ is the kinematic viscosity and $\eta$ is the magnetic diffusivity, so that
$Pr_M= Re_M/Re= \nu/\eta$.

Typical values of $Pr_M$, $Re_M$, and $Re$ for astrophysical systems have been compiled by Brandenburg \& Subramanian
(2005) using the microscopic (Spitzer)  values for both the magnetic resistivity
and the kinematic viscosity. In most of the cases $Re$ and $Re_M$ are  very large because of the large scales involved in astrophysical systems, and  $Pr_M$ is generally different from 1.
 For partially ionized gas, one finds that (e.g., Brandenburg \& Subramanian 2005)
 $Pr_M < 1$ in dense environments, such as stars (for which $Pr_M \sim 10^{-4}$) and  accretion disks.
 In these cases  $l_{\eta} > l_{\nu}$, where  $l_{\eta}$ corresponds to the scale at which the turbulent magnetic fields diffuse and $l_{\nu}$ corresponds to the scale where turbulence dissipates.
 While $Pr_M > 1$ in small density environments, such as galaxies ($Pr_M \sim 10^{14}$) and clusters of galaxies,
 implying  $l_{\eta} < l_{\nu}$.

The different regimes above will determine the scale at which an SSD saturates.
For instance, for a system with
$Pr_M \gg1$ and $Re \sim 1$, Schekochihin et al. (2004) have found that the SSD
 spreads most of the magnetic energy
over the sub-viscous range and piles it up at the magnetic resistive scales resulting in a very folded magnetic field structure. However, this does not seem to be the case  when $Re\gg1$.

For systems with
$Pr_M\gg1$ and $Re\gg1$ (typical of galaxies and clusters), numerical simulations indicate that both folded and  non-folded  magnetic field structures should coexist (Brandenburg \& Subramanian 2005).

 Consistent with the results shown in the previous paragraph, for systems with $Pr_M \sim 1$ (implying  $Re=Re_M$), numerical studies by Haugen et al. (2003, 2004) have shown that the  magnetic field  correlation lengths at the saturated state are of the order of 1/6 of the velocity correlation scales and therefore much larger than the magnetic  resistive scale.


 For systems with $Pr_M\ll1$ and $Re\gg1$ (as one expects in the case of stars and accretion disks), since
  $k_{\eta} \ll k_{\nu}$ most of the energy is dissipated resistively leaving very little kinetic energy to be cascaded and terminating the kinetic energy cascade earlier than in the case of a system with  $Pr_M = 1$.

\section{Small scale dynamos applied to the intra-cluster medium   }

In the previous section we presented a brief review of the current status of large and small scale turbulent dynamos theories in general.  In this section we will focus on the role of small scale turbulent dynamos in cluster of galaxies.

 Magnetic fields in the ICM  are observed to be turbulent (Ensslin \& Vogt 2005; Govoni et al. 2005; see also Hanasz압 chapter in these Proceedings).

As argued by Parker (1979), Brandenburg et al. (2012) and others, the cosmic magnetic fields detected in galaxies and in the ICM cannot have a primordial origin only, because in order to sustain these fields  against turbulent
decay,  dynamo action seems to be required. Otherwise, the Lorentz forces due to these fields would rapidly
transfer magnetic energy to kinetic energy which in turn would be dissipated by viscosity or magnetic reconnection diffusion  in the turbulent flow. Thus it becomes hard to make a convincing case for purely
primordial magnetic fields without any dynamo action to explain the observed cosmic magnetism (Brandenburg et al. 2012).

Only SSDs must operate in the ICM, as  emphasized in Section 1
  (Subramanian et al. 2005;   Schober압 chapter in these Proceedings). An SSD will amplify
 seed fields which are injected in the ICM by AGNs, galactic winds, and galaxy mergers.

A much less explored problem in the framework of turbulent dynamo action  in cluster of galaxies is the fact that the ICM is collisionless. Its low   densities ($10^{-3}$  to $10^{-2}$ cm$^{-3}$) imply  an
ion Larmor radius   much smaller than the      mean free path for   binary collisions. In the Hydra cluster, for instance, the ion Larmor radius is $\sim 10^5$ km, while the particles mean free path is $\sim 10^{15}$ km (Ensslin \& Voigt 2006). This makes the application of a standard (collisional) MHD formulation inappropriate in this case. A way to solve this problem is to apply a kinetic description for the ICM, however, such an approach is not appropriate either for studying the large scale phenomena in these environments and, in particular, the evolution of the turbulence and magnetic fields.

Fortunately, it is
possible to formulate a fluid approximation for collisionless plasmas, namely,  a collisionless-MHD approach.
The low rate of collisions in the fluid leads to anisotropy of the thermal pressure. In this case, it is possible to assume  a double Maxwellian velocity distribution of the particles in the directions parallel and perpendicular to the local magnetic field which result in distinct pressure terms in both directions. The simplest collisionless-MHD approximation that introduces  this  pressure anisotropy in the MHD formulation was first proposed  by Chew,  Goldberger \& Low (Chew et al. 1956), the so called CGL-MHD model.

On the other hand, the forces arising from this anisotropy in the MHD equations modify the standard Alfv\'en and magnetosonic waves and lead to the development of kinetic instabilities such as the mirror and the firehose instabilities (e.g., Kulsrud 1983). The mirror instability dominates when the thermal pressure component perpendicular to the local magnetic field is larger than the parallel component and it tends to accumulate gas in regions of smaller magnetic field. The firehose instability is dominant  in regions where the parallel component of the thermal pressure is larger and it tends to bend the
lines and trap gas in zones of larger magnetic field.

Measurements from weakly collisional plasmas, as those in the solar wind or in the magnetosheath,  and in laboratory experiments, as well as PIC numerical simulations have  demonstrated that these instabilities are able to redistribute the pitch angles of the particles, thus decreasing the  anisotropy (Gary 1993).
A modified  CGL-MHD model taking into account the  constraints on the anisotropy  due to the back reaction of these kinetic instabilities has been recently employed for modelling the solar wind
(Samsonov et al. 2007; Chandran et al. 2011).
Kunz
et al. (2011) have also employed a  collisionless-MHD approach including a semi-phenomenological model for heating the central regions of  galaxy clusters with cold cores, which is able to counterbalance the thermal emission losses, therefore
preventing the non observed cooling flows in these systems.  This heating is originated at the conversion of turbulent to thermal
energy by the micro-instabilities driven by the temperature anisotropy.

Using a similar collisionless-MHD model with phenomenological constraints on  the growth of the pressure/temperature anisotropy, we have explored  the evolution of the turbulence and the amplification of seed magnetic fields due to small scale turbulent dynamo action  in the collisionless plasma of the intracluster medium.
For this aim, we employed an one-fluid three-dimensional collisionless-MHD code, forcing non-helical turbulence into a periodic cubic box (see below and more details in Santos-Lima  et al. 2012c).

\subsection{ The collisionless-MHD equations}

The mass,  momentum, and  induction equations in the  collisionless-MHD approximation write (Santos-Lima et al. 2012c):

\begin{equation}
\frac{d \rho}{d t}  = - \rho \nabla \cdot \mathbf{u}
\end{equation}
\begin{equation}
\rho \frac{d \mathbf{u} }{d t} =
- \nabla_{\mathbf{t}} p_{\perp} - \nabla_{\mathbf{b}} p_{\parallel}
+ \frac{1}{4 \pi} \mathbf{\left( \nabla \times B \right) \times B}
- \frac{p_{\perp} - p_{\parallel}}{B} \nabla_{\mathbf{b}} B
\end{equation}
\begin{equation}
\frac{d \mathbf{B}}{d t} =  - \mathbf{B} (\nabla \cdot \mathbf{u}) + (\mathbf{B} \cdot \nabla) \mathbf{u}
\end{equation}

\noindent Where the variables have their usual definitions, $\nabla_{\mathbf{b}}$ is the gradient taken in the direction of the magnetic field, 
and $p_{\parallel}$ and $p_{\perp}$ are the thermal pressure components parallel and perpendicular to the magnetic field, respectively.

In the standard CGL-MHD model,  the pressure anisotropy $A=p_{\perp}/p_{\parallel}$, which can grow indefinitely with no constraints, is computed by assuming that the flow is adiabatic and that magnetic momentum conservation holds.
This implies the following closure (Chew et al. 1956):

\begin{equation}
\frac{d }{d t} \left( \frac{p_{\perp}}{\rho B} \right) = 0
\end{equation}
\begin{equation}
\frac{d }{d t} \left( \frac{p_{\parallel} B^{2}}{\rho^{3}} \right) = 0
\end{equation}

Recently, Kowal et al. (2011; see also Santos-Lima et al. 2011)  considered the CGL-MHD approximation with an isothermal closure in order to account for the radiative losses of the plasma. In this CGL-isothermal approach, the temperatures in the parallel and perpendicular directions to the local magnetic field are assumed to be constant.

Neither of the models above take into account the saturation of the pressure anisotropy growth due to the back reaction of the kinetic instabilities. In order to allow for their feedback, we have adopted different phenomenological approximations (Santos-Lima et al. 2012c). One of them includes a similar approach to that employed by Samsonov et al. (2001) for the  solar wind collisionless plasma. In this case, pressure isotropization  due to the kinetic instabilities  is applied whenever the region considered reaches an anisotropy  above a threshold $A^{\star}$:

\begin{equation}
\left( \frac{\partial p_{\perp}}{\partial t} \right) _{DIFF} = 
\begin{cases}
-p_{\perp} \nu_{DIFF} (A - A^{\star}_{mi}), &\mbox{if } A \ge A^{\star}_{mi} \\
-p_{\perp} \nu_{DIFF} (A - A^{\star}_{fh}), &\mbox{if } A \le A^{\star}_{fh} \\
0, &\mbox{otherwise}
\end{cases}
\end{equation}

\noindent Where $A^{\star}_{mi}$ corresponds to the anisotropy threshold due to the mirror instability back reaction  and $A^{\star}_{fh}$ corresponds to the anisotropy threshold due to the firehose instability back reaction on the plasma (numerically, these thresholds are given by the marginal values of both instabilities). $\nu_{DIFF}$ gives the isotropization rate which is of the order of the maximum  growth rate of the dominant kinetic instability (Santos-Lima et al. 2012c).

\subsection{ Small scale turbulent dynamo in a collisionless-MHD fluid applied to the ICM}

We have integrated numerically the set of equations described above, considering the different collisionless-MHD models, i.e., the standard CGL, 
the isothermal-CGL\footnote{
We should note that, distinctly from the double-isothermal-CGL model investigated by 
 Kowal et al. 2011 where each component of the thermal velocity was assumed to be constant,  in the present work we have assumed the total (parallel plus perpendicular) thermal velocity to be constant in the isothermal-CGL model.}, 
and the model with isotropization constraints. For comparison, we have also performed simulations considering a standard (collisional) MHD-model.
Figure 2 shows an example of such simulations. For a more extensive parametric study including also higher resolution simulations, we refer to Santos-Lima et al. 2012c).
In all the simulations shown in Figure 2, non-helical turbulence with a  velocity $1 c.u.$  is injected at
a scale 0.4 of the cubic box size.   The pressure is initially
isotropic. The initial  magnetic field seed is uniform with intensity $B_o= 10^{-4}$ c.u.

\begin{figure}
\centering
\resizebox{12.6cm}{!}{\includegraphics{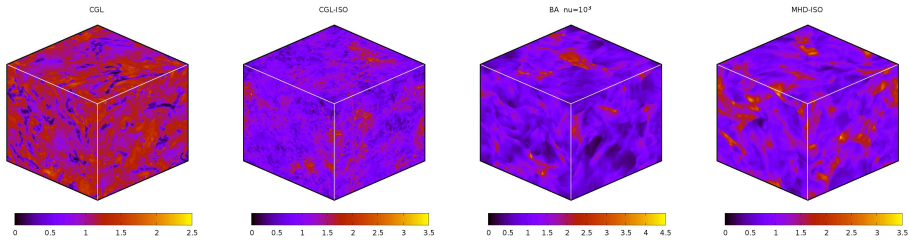} }
\caption[]{3D cubes showing the density distribution for different turbulent systems at t= 3 c.u. The initial seed magnetic field is
$B_o = 10^{-4 }$ c.u. and the sound speed is $c_s=1$ c.u. in all models (since the injected turbulent velocity $v=1$ c.u., the turbulence is transonic in all models depicted). From left to right: (a) CGL model; (b)isothermal-CGL; (c) CGL with isotropization (labelled BA); and (d) collisional MHD model.}
\label{fig2}
\end{figure}

The 3D collisionless models of Figure 2 mimic typical conditions of the ICM. In the CGL and isothermal-CGL models there is no constraint on the growth of the pressure anisotropy. In these cases, the
kinetic instabilities  that develop due to the anisotropic pressure are very strong at the smallest scales and accumulate most of the energy there,
making the density (and magnetic field) distribution much more ``wrinkled'' than in the standard (collisional MHD case).
On the other hand, in the model where the isotropization of the thermal pressure due to the back reaction of the same  kinetic instabilities is allowed to act above an isotropy threshold, the developed density (and magnetic field) structures are larger and more similar to those of the collisional MHD turbulent model.

\begin{figure}
\centering
\resizebox{6.3cm}{!}{\includegraphics{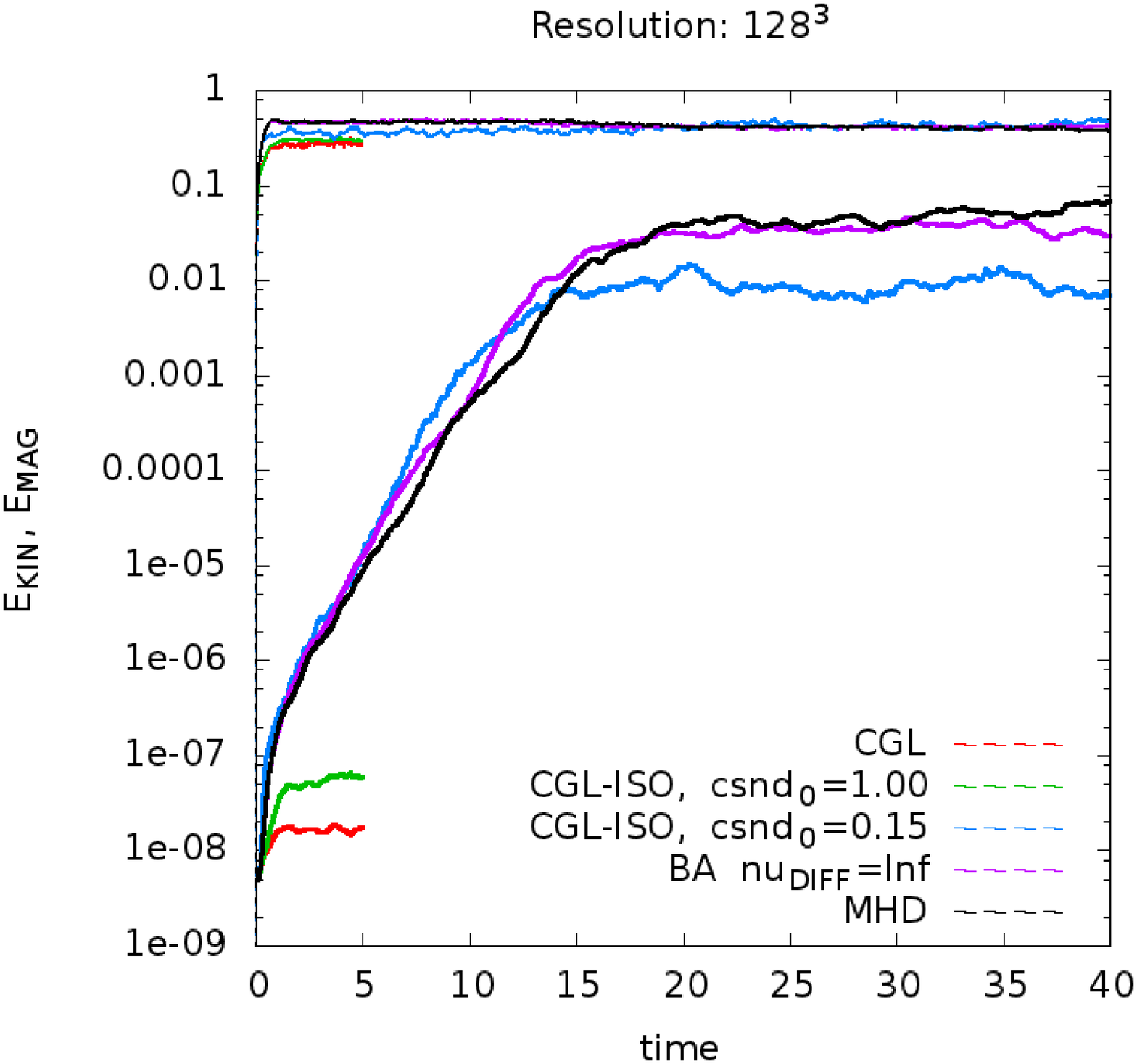} }
\resizebox{6.3cm}{!}{\includegraphics{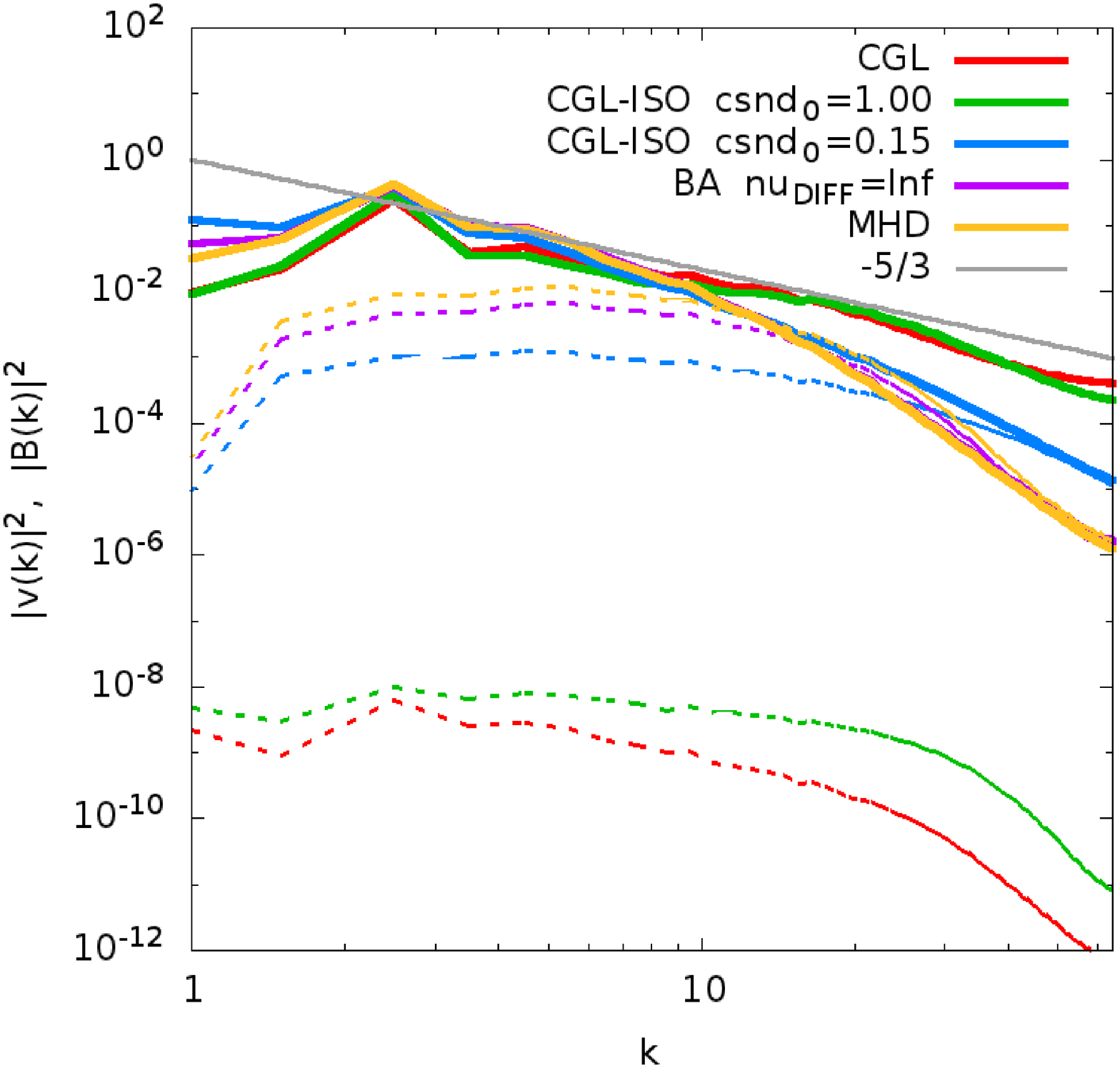} }
\caption[]{
Left: evolution of the kinetic and magnetic energies for the different
models depicted in Figure 2. An isothermal-CGL model with sound speed $c_s = 0.15$ c.u. is also included for comparison (in this case, the injected turbulence is supersonic). The model labelled BA corresponds to the CGL model with isotropization in Figure 2. Right: power spectrum of the velocity and magnetic fields averaged
over the time interval t= 35 to 40 c.u.   
}
\label{fig3}
\end{figure}

Figures 3a and 3b  depict  the magnetic energy growth as a function of time and the power spectrum of the magnetic and kinetic energies for the turbulent dynamos of Figure 2.
We clearly see in Figure 3b (right)
that the power spectra of the collisionless CGL and isothermal-CGL models are ``harder'' (or more intense) at the highest values of k, i.e., at the smallest scales of the system, than the MHD model. This confirms the trend of Figure 2, where we observe the accumulation of structures at the smallest scales in these models due to the  kinetic instabilities. On the other hand, the CGL model with the isotropization condition (the model labelled BA) has a power spectrum that evolves similarly to the collisional MHD model.

On the left diagram of Figure 3 (Figure 3a), we see the dynamo action of the turbulence with an initial exponential amplification of the seed field followed by a slower linear, almost saturated growth, at later times. In the case of the collisionless model with isotropization (BA model in the figure), the growth of the magnetic field up to values of 10$\%$ of the kinetic energy of the turbulence is similar to that of the MHD model. On the other hand, for the collisionless models without the isotropization closure (CGL and isothermal-CGL with transonic turbulence) there is an incipient amplification of the magnetic field. In other words, no dynamo operates when the turbulence is subsonic or transonic and pressure anisotropy is allowed to grow indefinitely  accumulating energy at the smallest scales.
This occurs because in these cases (with no isotropization condition) the
turbulence allows the continuous  increase of the pressure anisotropy ($A=p_{\perp}/p_{\parallel}$).
When   $A=p_{\perp}/p_{\parallel} >$ 1, the thermal energy is always dominant over the  magnetic energy and subsonic turbulence is unable to stretch or bend the field lines, so that  there is no dynamo amplification. In the
more realistic case, when the pressure isotropization is included, eventually  $A \rightarrow  1$ due to   the instabilities back reaction on the plasma and then, even for initial high $\beta = p_{th}/p_M$  and transonic or subsonic turbulence, the lines are stretched and bent and there is dynamo amplification, as we see in Figure 3a. In the absence of isotropization, Figure 3a also shows that a dynamo amplification is still possible in an isothermal-CGL model, but the turbulence has to be highly supersonic in this case. 
(The dynamo action is enhanced in this  case because the modified Alfv\'en velocity by the anisotropy remains smaller than the turbulent velocity; see  Santos-Lima et al. 2012c.)

The results above are very interesting and stimulating as they indicate that,  as long as  turbulence and dynamo amplification of seed magnetic fields are concerned, one can still treat the ICM as a nearly collisional MHD environment. This, of course, provided that the  pressure isotropization approach used above is consistent.
Based on  solar wind and magnetosheath observations and on laboratory experiments, as well as on PIC simulation results, the adoption of a pressure isotropization closure as just described seems to be appropriate. Nonetheless, further studies exploring better the microphysics of the kinetic instabilities for obtaining more self-consistent closures and values of the anisotropy thresholds are still needed.

\section{Turbulent diffusion and the role of fast magnetic reconnection in turbulent dynamos }

Richardson압 diffusion in turbulent flows (1926) indicate that
the particles suffer spontaneous stochasticity, as a consequence there is an
explosive separation into larger and larger turbulent
eddies that cause an efficient turbulent diffusion in the flow. An important implication of this result is   that
magnetic flux conservation in turbulent fluids is violated! It is only stochastically conserved, as claimed by  Eyink (2011).

On the other hand, Lazarian \& Vishniac (1999) had long explored the effects of turbulence on magnetic reconnection and found that when turbulence is present, reconnection is $fast$.

Magnetic reconnection occurs when two magnetic fluxes of opposite polarity encounter each other.
In the presence of finite magnetic resistivity, the converging magnetic lines annihilate at the discontinuity surface and a current sheet forms.
In the standard Sweet-Parker (S-P) model of magnetic reconnection, the velocity at which  two converging magnetic fluxes of opposite polarity reconnect is given by $v_{rec} \approx v_A S^{1/2}$, where
$S = l v_A /\eta$,  $\eta$ is the Ohmic diffusivity, $l$ is a typical  scale of the system and $v_A$ is the Alfv\'en velocity. For astrophysical systems $l$ is in general very large and therefore, $S\gg 1$.
  Because $S$ is large for Ohmic resistivity values, the S-P reconnection is very slow.
However, Lazarian \& Vishniac (1999) have demonstrated that  when turbulence is present, the
magnetic field wandering at all turbulent scales within the current sheet allows the formation of a thick volume filled with several reconnected small magnetic fluctuations which make the reconnection fast. This model was successfully tested numerically by Kowal et al. (2009, 2012) and has challenged the well-rooted concept of magnetic field frozenness for the case of turbulent fluids.

A natural consequence of the fast reconnection in turbulent flows is that it  provides an efficient way by which  magnetic flux can diffuse through the turbulent eddies in  astrophysical flows, particularly when the turbulence is super-Alfv\'enic. The theoretical grounds of this  ``reconnection diffusion''  mechanism in turbulent flows have been  described in detail in several recent reviews  (Lazarian 2005; Lazarian 2011; Lazarian et al. 2011;  de Gouveia Dal Pino et al. 2011; 2012; Eyink et al. 2011). Also, it has been successfully tested numerically in the context of star formation and molecular clouds by Santos-Lima et al. (2010, 2012a, 2012b) and Le\~ao et al. (2012).

Figure 4 shows a schematic representation on how interacting turbulent eddies can mix the gas  and exchange parts of their magnetic flux tubes (through reconnection) favouring their diffusion.
This theory predicts a turbulent reconnection diffusivity $\eta_t$ which is much larger than the Ohmic diffusivity at the turbulent scales:
 (Lazarian 2005; Santos-Lima et al. 2010; Lazarian 2006; 2011; Lazarian et al. 2012; Le\~ao et al. 2012):

\begin{equation}
\begin{split}
&\eta_t \sim l_{\rm inj}v_{\rm turb} \ \  \  \  \  \  \ \ \ \ \ \ \ \ \;  {\rm if}\;  v_{\rm turb} \geq v_A\; ,\\
&\eta_t \sim l_{\rm inj}v_{\rm turb} \left(\frac{v_{\rm turb}}{v_A}\right)^3\; \; {\rm if}\; v_{\rm turb} < v_A\; ,
\end{split}
\label{diffusivity}
\end{equation}

\noindent  where $l_{inj}=L/k_f$  and $v_{\rm turb}=v_{rms}$. The relations above indicate that the ratio $\left(v_{\rm turb}/v_A\right)^3\;$ is important only in a regime of  sub-Alfv\'enic turbulence, i.e. with the  Alfv\'enic Mach number  $M_A \leq 1$. We also notice that when  $v_{\rm turb} \geq v_A$ the predicted diffusivity is similar to the Richardson's  turbulent diffusion coefficient, as one should expect.


\begin{figure}
\centering
\resizebox{6.3cm}{!}{\includegraphics{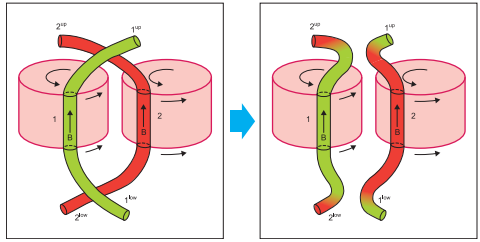} }
\caption{Schematic representation of two interacting turbulent eddies each one carrying its own magnetic flux tube. The turbulent interaction causes an efficient mixing of the gas of the two eddies as well as fast magnetic reconnection of the two flux tubes which leads to  diffusion of the magnetic field (extracted from Lazarian 2011). }
\label{fig4}
\end{figure}

Figure 13 of Le\~ao et al. (2012) presents a comprehensive  example from a 3D MHD numerical simulation that shows  how effective turbulent reconnection diffusion is  at removing the excess of magnetic flux from the central regions of a turbulent molecular cloud allowing the gravitational collapse of its central core to form stars.


Likewise, one may question how this universal mechanism of reconnection diffusion can affect turbulent dynamos?
We know that, specially in  large scale turbulent dynamos, i.e., within the mean field dynamo theory,  an efficient dissipation of the small scale magnetic fields that develop during the dynamo action is a crucial step in order to allow for the development and survival of the large scale magnetic fields. This dissipation is commonly attributed to the turbulent (Richardson) diffusion which is approximately equal to the first equation in \ref{diffusivity} above. On the other hand,
 the examination of the second equation in \ref{diffusivity}, shows that when the Alfv\'en velocity becomes dominant over the turbulent velocity in the process of magnetic field amplification at a given scale,  the effective diffusion in this regime becomes much smaller than the standard Richardson's turbulent diffusion. This may inhibit the dissipation of the small scale magnetic fields at the saturation regime of the dynamo and thus prevent the growth of the large scale fields. This possibility  should be  examined through  numerical studies. The effects of fast magnetic reconnection  on the magnetic flux pumping in large scale dynamos processes (e.g. Guerrero \& de Gouveia Dal Pino 2008)  should be  explored as well (see, e.g., de Gouveia Dal Pino et al. 2012).

\section{Summary and Conclusions }

In this lecture we have discussed some interlinks between turbulence and dynamo processes. After a  very brief summary  of the current status of turbulent dynamo theories,  will focussed  on small scale turbulent dynamos. These are believed to be particularly important to explain the  amplification and maintenance  of magnetic fields  in the  intra-cluster media (ICM) of galaxies.

However, the collisionless nature of the fluid in the ICM puts in question the applicability of
 standard fluid theories (which assume isotropic thermal pressures) to study its plasma dynamics.  A $collisionless-MHD$  approach seems to be more appropriate to describe large scale phenomena, such as turbulence and magnetic field dynamo amplification in these environments.
In this case, one can assume a double Maxwellian velocity distribution of the flow in both directions, parallel and perpendicular to the local magnetic field, which gives rise to an anisotropic thermal pressure. The forces arising from this anisotropy lead to the development of kinetic instabilities.
Measurements from weakly collisional plasmas in the solar wind and in laboratory experiments, as well as PIC simulations have demonstrated in turn, that these instabilities
 redistribute the pitch angles of the particles, thus decreasing the pressure anisotropy.

Employing  a collisionless-MHD model with  phenomenological constraints on the pressure anisotropy due to the back reaction of these kinetic instabilities, we have studied numerically the  turbulent amplification of seed magnetic fields  in the collisionless plasma of the ICM. The comparison of  the  results  of  this collisionless-MHD model   with those of a standard (collisional) MHD-model with similar initial conditions have revealed  that the magnitude of amplification and the resulting distribution of the magnetic field as well as, the  magnetic energy  power spectrum  are similar in both models.
These results  are particularly  interesting as they suggest  that,  as long as  turbulence and dynamo amplification of seed magnetic fields are concerned, one can still treat the ICM as a nearly collisional MHD environment. This, of course, provided that the  pressure isotropization closure used here is appropriate to the ICM (see more details in Santos-Lima et al. 2012c). Further studies in this regard are still required in order to build a self-consistent model of the pressure isotropization rate due to the back reaction of the kinetic instabilities which are triggered by the anisotropy itself. PIC simulations may be needed in this case.

Finally, we have also addressed briefly the role that turbulent magnetic reconnection diffusion may have on large scale dynamos action.
It has been shown that the presence of turbulence makes magnetic reconnection fast and this in turn can make  the diffusion of magnetic flux very efficient, particularly in the regime of super-Alfv\'enic turbulence. In this case, the reconnection diffusivity  is of the same order of the Richardson's turbulent diffusivity. However, in a regime of sub-Alfv\'enic turbulence, the reconnection diffusivity decreases by a factor $(v_A/v_{turb})^3$. This may have important implications for a large scale turbulent dynamo action.
We know that in this  case,  an efficient dissipation of the small scale magnetic fields that develop during the dynamo action is a crucial step in order to allow for the development of the large scale magnetic fields.  However, according to the discussion above, when the Alfv\'en velocity becomes dominant over the turbulent velocity in the process of magnetic field amplification at a given scale,  the effective diffusion in this region becomes much smaller than the standard Richardson's turbulent diffusion. This may inhibit the dissipation of the small scale magnetic fields at the saturation regime of the dynamo and thus prevent the growth of the large scale fields. This possibility still requires   numerical investigation.

\section{Acknowledgements}
The authors are in debt with 
Gustavo Guerrero 
and Alex Lazarian for very fruitful and stimulating discussions during the production of this short review. 
 E.M.G.D.P also acknowledges support from the Brazilian agencies  FAPESP (2006/50654-3) and CNPq (300083/94-7), and R.S.L. acknowledges support from  FAPESP (2007/04551-0).   Part of the numerical simulations here presented were performed  in the supercomputer Alfa-Crucis of the Astrophysical Informatics Laboratory LAi of IAG, Astronomy Department, University of S伋 Paulo (funded by FAPESP: 2009/54006-4).

%
%
%

\end{document}